\documentclass[apjl]{latex/emulateapj}

\usepackage{epsfig}
\usepackage{xspace}
\usepackage{amsmath}
\usepackage{color}

\slugcomment{Submitted to ApJ}

\shorttitle{Deconstructing Thermal SZ -- Lensing Cross-Correlations}
\shortauthors{Battaglia, Hill, \& Murray}

\newcommand{\be}{\begin{equation}}
\newcommand{\ee}{\end{equation}}
\newcommand{\bea}{\begin{eqnarray}}
\newcommand{\eea}{\end{eqnarray}}
\newcommand{\rmn}{\mathrm}

\newcommand{\st}{\sigma_\mathrm{T}}
\newcommand{\me}{m_\mathrm{e}}
\newcommand{\nume}{n_\mathrm{e}}
\newcommand{\dd}{\mathrm{d}}
\newcommand{\Sig}{\kappa_i}

\newcommand{\Wcmb}{W_\mathrm{CMB}}
\newcommand{\Wgal}{W_\mathrm{gal}}
\newcommand{\yphiname}{y \otimes \phi_\mathrm{CMB}}
\newcommand{\ygalname}{y \otimes \phi_\mathrm{GAL}}
\newcommand{\yphi}{C_\ell^{\phi y}}
\newcommand{\ygal}{C_\ell^{\phi_g y}}

\begin{document}
\title{Deconstructing Thermal Sunyaev-Zel'dovich -- Gravitational Lensing Cross-Correlations: Implications for the Intracluster Medium}

\author{N. Battaglia$^{1,2}$, J. C. Hill$^{3,1}$, N. Murray$^{4}$}

\altaffiltext{1}{Department of Astrophysical Sciences, Princeton University, Princeton, NJ 08544}
\altaffiltext{2}{McWilliams Center for Cosmology, Carnegie Mellon University, 5000 Forbes Ave, Pittsburgh PA, USA, 15213}
\altaffiltext{3}{Department of Astronomy, Columbia University, New York, NY 10027}
\altaffiltext{4}{Canadian Institute for Theoretical Astrophysics, 60 St George, Toronto, ON M5S 3H8, Canada}

\begin{abstract}

Recent first detections of the cross-correlation of the thermal Sunyaev-Zel'dovich (tSZ) signal in \emph{Planck} cosmic microwave background (CMB) temperature maps with gravitational lensing maps inferred from the \emph{Planck} CMB data and the \emph{CFHTLenS} galaxy survey provide new probes of the relationship between baryons and dark matter. Using cosmological hydrodynamics simulations, we show that these cross-correlation signals are dominated by contributions from hot gas in the intracluster medium (ICM), rather than diffuse, unbound gas located beyond the virial radius (the ``missing baryons''). Thus, these cross-correlations offer a tool with which to study the ICM over a wide range of halo masses and redshifts. In particular, we show that the tSZ -- CMB lensing cross-correlation is more sensitive to gas in lower-mass, higher-redshift halos and gas at larger cluster-centric radii than the tSZ -- galaxy lensing cross-correlation. Combining these measurements with primary CMB data will constrain feedback models through their signatures in the ICM pressure profile. We forecast the ability of ongoing and future experiments to constrain such ICM parameters, including the mean amplitude of the pressure -- mass relation, the redshift evolution of this amplitude, and the mean outer logarithmic slope of the pressure profile. The results are promising, with $\approx 5-20$\% precision constraints achievable with upcoming experiments, even after marginalizing over cosmological parameters.

\end{abstract}

\keywords{Cosmic Microwave Background --- Cosmology: Theory ---
  Galaxies: Clusters: General --- Large-Scale Structure of Universe
   --- Methods: Numerical}

\section{Introduction}

Modeling the thermodynamic and dark matter (DM) properties of halos as structure grows in the Universe is fundamental to our understanding of the physics involved in galaxy formation and cosmology. In a simple model for cosmological structure formation, the thermal properties of the gas in massive halos ($\sim 10^{13} - 10^{15}\,M_\odot$), known as the intracluster medium (ICM), are determined by the DM-dominated gravitational potential through spherical collapse \citep{Kaiser1986}.  Such a model predicts self-similar scalings of the global thermodynamic properties of halos as a function of their mass and redshift. Invoking equilibrium and symmetry arguments along with the shape of the gravitational potential, one can extend this model to predict radial ICM profiles, such as the entropy profile \citep[e.g.,][]{Voit2002,CPL2009} or pressure profile \citep[e.g.,][]{KS2001,OBB2005}. However, observations \citep[e.g.,][]{Horner2001,Vik2006} provide significant evidence that non-thermal processes such as star formation, radiative cooling, turbulence, and feedback contribute to the energetics of the ICM. In cosmological hydrodynamic simulations, these processes are modeled with {\it sub-grid} methods \citep[e.g.,][]{2000ApJ...536..623L,2003MNRAS.339..289S,2006ApJ...650..538N,Sijacki2007,BS2009,Dub2012} and calibrated to measurements of halo properties at low redshifts. Proper calibration of these sub-grid models requires observables that are sensitive to the thermodynamic properties across decades in halo mass and out to high redshift.

Secondary anisotropies of the cosmic microwave background (CMB) contain an abundance of cosmological and astrophysical information at $z \lesssim 10$. Due to advances in resolution and sensitivity achieved by recent CMB experiments, such as the Atacama Cosmology Telescope \citep[ACT/ACTPol,][]{ACT,ACTpol}, the South Pole Telescope \citep[SPT/SPTPol,][]{SPT,SPTpol}, the \emph{Planck} satellite \citep{PLNK}, and {\sc Polarbear} \citep{PolarB}, it is now possible to extract this information. The secondary anisotropies of interest in this work are those sourced by the thermal Sunyaev-Zel'dovich (tSZ) effect and gravitational lensing.

The tSZ effect is the Compton up-scattering of CMB photons by hot electrons, leading to a unique spectral distortion in the CMB that is negative at frequencies below $\approx 220$ GHz and positive at higher frequencies \citep{SZ1970}.  The amplitude of this distortion, sometimes known as the ``Compton-$y$'' signal, is proportional to the electron pressure integrated along the line of sight.  As a result, the largest tSZ signals arise from electrons in the ICM of massive galaxy clusters. Several hundred new massive clusters have been detected in blind mm-wave surveys via the tSZ effect \citep[e.g.,][]{Hass2013,PlanckClustCat,Bleem2014}, and the tSZ signal has now been observed at lower mass scales through stacking microwave maps on the locations of groups and massive galaxies \citep[e.g.,][]{Hand2011,PlankStkLBG,Greco2014}. The tSZ effect has also been measured statistically in the power spectrum \citep[e.g.][]{Dunk2011,Reic2012,Siev2013,PlnkY2013,George2014}, bispectrum or skewness \citep{Wilson2012,Crawford2014}, and the temperature histogram \citep{Hill2014}. However, uncertainties in ICM modeling limit the ability to use these statistical measurements to constrain cosmological parameters \citep[e.g.,][]{HP2013,McCarthy2014}. For example, at angular scales of $\ell = 3000$ half the power in the tSZ auto-spectrum comes from low-mass halos ($M \lesssim 2 \times 10^{14} M_\sun$) and high redshifts ($z \gtrsim 0.5$) \citep[e.g.,][]{2011ApJ...727...94T,BBPS2}. Additional uncertainties arise due to the modeling of other secondaries, such as the kinetic SZ effect, cosmic infrared background (CIB), radio sources, and the correlation between the CIB and tSZ signals.

The CMB lensing signal originates from the deflection of CMB photons by the gravitational field of matter located between the surface of last scattering and our telescopes. These deflections are small coherent distortions of roughly degree-scale CMB patches by $\approx 2-3$ arcminutes. It is possible to reconstruct the lensing potential from the statistical anisotropy induced by lensing in the small-scale power spectrum \citep[e.g.,][]{OH2003}. Similar to the recent advances in tSZ observations, CMB lensing has experienced a rapid growth from the first detections in cross- \citep{Smith2007,Hira2008} and auto-correlation \citep{Das2011,AvE2012} to the full-sky reconstruction of the lensing potential by \emph{Planck} \citep{PLNKLens}, as well as the first detections of polarization lensing \citep{Hans2013,PBC2013,PBC2014,AvE2014,SPTpol_Lens}. The CMB lensing signal is a robust tracer of the large-scale matter density field.  Thus, it correlates with a variety of halo populations over a wide redshift range \citep[e.g.][]{Sherwin2012,Bleem2012,Holder2013,PLNKIRLens}.

In addition to lensing of the CMB, weak gravitational lensing of light from background galaxies provides another tracer of the underlying matter density field \citep[e.g.,][]{Tyson1984,Kaiser1992}. The galaxy weak lensing signal appears as small but coherent distortions (``shear'') in galaxy shapes resulting from the gravitational deflection of light by intervening lenses along the line of sight. Matter over-densities produce tangentially oriented shear correlations. From the measured shear field, one can reconstruct a map of the lensing convergence. For a thorough review of weak lensing theory and observations, see \citet{BS2001}. Weak lensing is now the focus of a number of current and future galaxy surveys aiming to constrain the nature of dark energy \citep[e.g.,][]{CFHTdata2013,DES2005,HSC2012,LSST2009,Euclid2011,WFIRST}. Due to the different depths and galaxy populations probed, these surveys are sensitive to cosmic structure over different redshift ranges. Thus, cross-correlating other tracers with the different lensing convergence maps allows for tomography, an idea that we take advantage of below.

In this paper, we explore cross-correlations of the tSZ signal, which probes the ICM, and the weak lensing signals from the CMB and galaxies, which probe the matter distribution. We generally work with cross-power spectra in Fourier space, alleviating the effect of correlated errors present in real-space cross-correlation functions.  Early work on this topic focused on signal-to-noise estimates for then-upcoming CMB experiments such as WMAP and \emph{Planck} using simple theoretical models \citep{Goldberg-Spergel1999,Cooray-Hu2000,Coorayetal2000,Cooray2000}.  Recently, two $\approx 6\sigma$ measurements of tSZ -- lensing cross-correlations have been presented using the CMB data that was forecasted in the early studies. \citet{HS2014} constructed a Compton-$y$ map from the public \emph{Planck} data \citep{PlanckDATA} and cross-correlated it with the public CMB lensing potential map from \emph{Planck} \citep{PLNKLens}.  Interpreting the measured cross-power spectrum using analytic halo model calculations, they placed competitive constraints on the cosmological parameters $\sigma_8$ and $\Omega_\rmn{M}$ (assuming a fixed ICM physics model) and constraints on the ICM model (for a fixed background cosmology, with consistent results assuming either a WMAP9 or \emph{Planck} best-fit cosmology).  An independent Compton-$y$ map was constructed from the public \emph{Planck} data \citep{PlanckDATA} by \citet{vWHM2014}, who measured its cross-correlation with galaxy lensing shear maps from the Canada-France-Hawaii Telescope Legacy Survey ({\it CFHTLenS}) \citep{CFHTdata2013,CHFTMass2013}. From the measured real-space correlation function, \citet{vWHM2014} placed constraints on the gas fraction outside of halos. In a follow-up analysis, \citet{Ma2014} interpreted the same measurement using halo model calculations to claim the detection of a gas pressure profile in disagreement with that seen in X-ray observations of massive galaxy clusters at $z \lesssim 0.3$ \citep{Arnd2010}.

We re-examine the interpretation of both cross-correlation measurements in this paper using the cosmological hydrodynamics simulations described in \cite{BBPSS}.  Furthermore, we assess the validity of the analytic halo model calculations used previously to interpret the measurements by comparing them to the simulations.  Previous theoretical work on this topic focused on statistical moments \citep{Muns2014} and tomography \citep{Pratn2014}.  Here we focus specifically on the predictions of different ICM models for the tSZ -- lensing cross-correlations, while self-consistently considering the influence of cosmological parameter variations.  The interpretation of these cross-correlations in terms of ICM physics has important implications for understanding the discrepancy between cosmological parameters inferred from tSZ statistics \citep[e.g.,][]{Siev2013,PlnkY2013,George2014,Hill2014,McCarthy2014} and from the primordial CMB anisotropies \citep[e.g.,][]{PlanckParams}.  Moreover, in addition to re-interpreting the results of \cite{HS2014} and \cite{vWHM2014}, we also look ahead to upcoming measurements.

The capability to cross-correlate large areas of sky with high-quality lensing and CMB data will soon be possible. Near-future high-resolution CMB experiments on the ground, such as AdvACT \citep[e.g.][]{Calabrese2014} and SPT-3G \citep{SPT3G} will provide higher signal-to-noise multi-frequency maps across large areas of sky, which should further improve the signal to noise in future Compton-$y$ maps. When forecasting future measurements, we assume that the signal-to-noise of the Compton-$y$ map will improve by a factor of $\sqrt{5/2}$, representing the raw increase in data volume from the \emph{Planck} nominal mission data \citep{PLNK} used in \cite{HS2014} and \cite{vWHM2014} to the final \emph{Planck} results. CMB lensing reconstruction will also improve substantially with upcoming experiments --- for example, AdvACT should detect the CMB lensing power spectrum at signal-to-noise $\gg 100$. Galaxy lensing advances will be made over the pioneering work of the {\it CFHTLenS} survey. Ongoing experiments include the Dark Energy Survey \citep[DES,][]{DES2005} and Hyper Suprime Cam \citep[HSC,][]{HSC2012} imaging surveys, which will cover more area and image fainter galaxies than {\it CFHTLenS}. In the next decade, experiments such as the Large Synoptic Survey Telescope \citep[LSST,][]{LSST2009}, the \emph{Euclid} satellite \citep{Euclid2011}, and the Wide-Field InfraRed Survey Telescope \citep[WFIRST,][]{WFIRST} will provide further increases in sky area and signal-to-noise. Looking ahead to the shear maps from these surveys, the signal-to-noise of the tSZ -- galaxy lensing cross-correlations will be immense. Understanding these measurements will require further theoretical modeling of the gas and mass distributions in halos. 

This paper is organized as follows. In Section \ref{sec:meth}, we describe theoretical models for tSZ -- lensing cross-correlations, using both analytic calculations and numerical simulations. Section \ref{sec:res} compares simulations and analytic calculations of the cross-spectra and deconstructs these signals as a function of ICM physics model, halo mass, redshift, and cluster-centric radius. In Section \ref{sec:obscomp}, we compare the simulations and analytic results to measurements of the tSZ -- lensing cross-correlations. Section \ref{sec:fore} forecasts the constraints on ICM and cosmological parameters from future experiments. We present our conclusions in Section \ref{sec:conc}. 

We adopt a flat $\Lambda$CDM cosmology throughout. Note all masses quoted in this work are given relative to $h = 0.7$, where $H_0=100\,h\,\rmn{km}\,\rmn{s}^{-1}\,\rmn{Mpc}^{-1}$, unless stated otherwise. For compactness, we denote the tSZ -- CMB lensing cross-correlation as $\yphiname$ and the tSZ -- galaxy lensing cross-correlation as $\ygalname$.

\section{Methodology}
\label{sec:meth}

The cross-correlation between the tSZ effect and weak lensing probes the relationship between hot, ionized gas and gravitational potential. The signal strength of the tSZ spectral distortion in the observed CMB temperature is a function of frequency $\nu$ and the Compton-$y$ parameter:

\be
\frac{\Delta T(\nu)}{T_\rmn{CMB}} = f(\nu) y, 
\ee

\noindent where $f(\nu) = x\,\rmn{coth}(x/2) - 4$, $x = h\nu / (k_\rmn{B} T_\rmn{CMB})$, $k_\rmn{B}$ is Boltzmann's constant, and $T_\rmn{CMB}$ is the CMB temperature. We neglect relativistic corrections to the tSZ spectral function $f(\nu)$ \citep[e.g.,][]{Nozawaetal2006}, as the tSZ -- lensing cross-correlations are dominated by halos for which these corrections are negligible (see Section \ref{sec:mz}). The magnitude of $y$ is a function of the integrated electron pressure along the line of sight:

\be
y = \frac{\st}{\me c^2} \int \nume k_\rmn{B}T \dd l
\label{eq:y}
\ee

\noindent where $\st$ is the Thomson scattering cross-section, $\me$ is the electron mass, $c$ is the speed of light, $\nume$ is the number density of free electrons, $l$ is the physical line of sight distance, and $T \equiv T_\rmn{e} - T_\rmn{CMB}$. Here the temperature of the free electrons, $T_\rmn{e}$, is much greater than the CMB temperature, $T_\rmn{CMB}$, so $T \simeq T_\rmn{e}$. For an ideal gas, $P_\rmn{e} = T_\rmn{e} k_\rmn{B} \nume$, so $y \propto \int P_e\dd l$. For a fully ionized and ion-equilibrated plasma, the integrated $y$ parameter probes the total thermal energy in a halo. Thus, measurements of $y$ are essential to understanding the thermodynamic properties of the baryons inside halos.

As photons travel toward an observer, their path is bent by the gravitational field sourced by matter along the line of sight. If these deflections are in the weak-field regime, this effect is known as weak gravitational lensing. To calculate this weak lensing signal we use the thin lens limit, where the thickness of the gravitational lens is much smaller than both the distances between the observer and lens and the lens and background source (CMB or galaxies). We parameterize the weak lensing signal by the lensing convergence $\Sig$, where $i$ denotes the choice of background photon field (i.e., the CMB or galaxies). The convergence is a function of the projected mass along the line of sight and a lensing kernel,

\be
\Sig =  \int W_i (z) \left( \rho - \bar{\rho}(z) \right) \dd l
\label{eq:sig}
\ee

\noindent where $\rho$ is the physical matter density (DM, gas, and stars), $\bar{\rho}(z) = \bar{\rho}(z=0) (1+z)^3$ is the mean physical matter density at redshift $z$, and $W_i$ is the lensing kernel. For galaxy lensing, the kernel is (in physical units)

\be
\Wgal (z) = \frac{4 \pi G \chi (z)}{c^2 (1 + z)} \int^\infty_z \dd z_s \, p_s (z_s) \frac{\left( \chi (z_s) - \chi (z) \right)}{\chi (z_s)},
\label{eq:wgal}
\ee

\noindent where $G$ is the gravitational constant, $p_s(z_s)$ is the redshift distribution of source galaxies (normalized to have unit integral), and $\chi (z)$ is the comoving distance to redshift $z$. The properties of this kernel depend on the imaging survey under consideration. For the completed \emph{CFHTLenS} survey, we use the $p_s(z)$ shown in Fig.~1 of \citet{vWHM2014}. For surveys where observations are ongoing or have not started, we estimate $p_s(z)$ as

\be
p_s(z) = \frac{z^2}{2z_0^3} e^{-z/z_0},
\ee

\noindent where $z_0 = 1/3$ for HSC, DES, and LSST (we refer to these surveys as ``HSC-like'' in figures). For \emph{Euclid} we choose $p_s(z)$ such that it matches \emph{CFHTLenS} \citep{Euclid2011}. Thus, we have both low- and high-redshift lensing surveys when combining measurements in the forecasts presented in Sec.~\ref{sec:fore}.

The CMB lensing kernel is a special case of Eq.~\ref{eq:wgal} in which the source distribution is replaced by a single source at $z_* \approx 1100$, i.e., $p_s(z) = \delta^D(z-z_*)$, where $\delta^D$ is the Dirac delta function. Thus, the kernel simplifies to

\be
\Wcmb (z) = \frac{4 \pi G \chi (z) \left( \chi_* - \chi (z) \right) }{c^2 \chi_* (1 + z)},
\label{eq:wcmb}
\ee

\noindent where $\chi_* = \chi (z_*)$. This kernel peaks at $z \approx 2$ and thus it probes higher redshift halos than those probed by any of the galaxy lensing kernels.

Lensing quantities can be equivalently represented via the lensing potential $\phi$, which is related to the lensing convergence through the relation

\be
\kappa(\hat{n}) = -\nabla^2 \phi(\hat{n}) / 2,
\ee

\noindent where $\hat{n}$ is line of sight unit vector and $\nabla$ is the two-dimensional Laplacian in the plane of the sky. We choose to work in terms of $\phi_i$ in our calculations, converting from $\Sig$ to $\phi_i$ in multipole space where the conversion is trivial, $\phi_{i,\ell} = 2 \kappa_{i,\ell} / (\ell(\ell +1))$.

\subsection{Analytic halo model calculations}
\label{sec:thry}

For the analytic calculation of the angular power spectrum of $\yphiname$ and $\ygalname$, we use the halo model formalism \citep[e.g.,][]{1988MNRAS.233..637C}, as is standard for such calculations \citep[e.g.,][]{KS2002,HP2013,HS2014,Ma2014}.  As shown in \citet{HS2014}, the total cross-power spectrum ($C^i_{\ell}$) has contributions from both the one-halo ($C^i_{\ell,\rmn{1h}}$) and two-halo ($C^i_{\ell,\rmn{2h}}$) terms,

\be
C^i_{\ell} = C^i_{\ell,\rmn{1h}} + C^i_{\ell,\rmn{2h}},
\ee

\noindent where $i$ refers to the lensing field considered in the cross-correlation. We denote the cross-power spectrum for $\yphiname$ as $C_{\ell}^{\phi y}$ and that for $\ygalname$ as $C_{\ell}^{\phi_g y}$. 

The $C_{\ell,\rmn{1h}}$ term is modeled as a randomly distributed Poisson process on the sky. In the flat-sky limit,

\be
C^i_{\ell,\rmn{1h}} = \int \dd z \frac{\dd V}{\dd z} \int \dd M \frac{\dd n}{\dd M} \tilde{y}_\ell (M,z) \tilde{\phi}_{i,\ell} (M,z),
\label{eq:1hterm}
\ee

\noindent where $\dd V/\dd z$ is the comoving volume per steradian, ${\dd n / \dd M}$ is the halo mass function, and $\tilde{y}_\ell (M,z)$ and $\tilde{\phi}_{i,\ell} (M,z)$ are the two-dimensional Fourier transforms of the Compton-$y$ and lensing convergence profiles, respectively. The mass $M$ in Eq.~\ref{eq:1hterm} is the virial mass as defined in \citet{BN1998}. The mass function used is from \citet{Tink2008} and the details of the calculations can be found in \citet{HP2013} and \citet{HS2014}. The convergence profile and conversions between mass definitions are calculated assuming an NFW density profile \citep{NFW1997} and the concentration-mass relation from \citet{2008MNRAS.390L..64D}. For the Compton-$y$ profile we use a parametrized pressure profile fit to the {\it AGN feedback} simulations described below. Full details of the fit can be found in \citet{BBPS2}. The profile is given by

\be
\frac{P}{P_{200}} =  \Pi_0
   \left(x/x_{\rmn{c}}\right)^{\gamma}\left[1 +
   \left(x/x_{\rmn{c}}\right)^{\alpha} \right]^{-\tilde{\beta}}, \ x \equiv
   r/R_{200},
\label{eq:prof}
\ee

\noindent where $\gamma = 0.3$, $\alpha = 1.0$, $\Pi_0$, $x_{\rmn{c}}$, and $\tilde{\beta}$ are parameters with power-law dependences on mass and redshift, and $P_{\Delta}$ is the self-similar amplitude for pressure at $R_{\Delta}$ \citep{Kaiser1986,Voit2005}:

\be
P_{\Delta} = \frac{G M_{\Delta} \rho_\rmn{cr}(z) \Omega_b \Delta}{2\, \Omega_\rmn{M} R_{\Delta}} \,.
\label{eq:PDelta}
\ee

\noindent Here, $R_{\Delta}$ is the cluster-centric radius enclosing a mass $M_{\Delta}$ such that the mean enclosed density is $\Delta$ times the critical density at the cluster redshift, $\rho_\rmn{cr}(z) \equiv 3 H_0^2 \left( \Omega_\rmn{M}(1+z)^3 + \Omega_{\Lambda} \right) / (8\pi\,G)$, where $\Omega_\rmn{M}$, $\Omega_{\Lambda}$, and $\Omega_b$ are the fractions of the critical density today in matter, vacuum energy, and baryons, respectively.

Later in the paper, we allow for freedom in the gas physics model by letting the normalized amplitude $P_0$ and power-law redshift dependence $\alpha_{z,P_0}$ of $\Pi_0$ vary, i.e.,

\be
\label{eq:P0def}
\begin{aligned}
\Pi_0(M_{200},z) & = 18.1 P_0 \\
	& \left( \frac{M_{200}}{10^{14} \,\, M_{\odot}} \right)^{0.154} \left( 1+z \right)^{\alpha_{z,P_0}} \,,
\end{aligned}
\ee

\noindent where the specific numbers are from the fitting function presented in \citet{BBPS2}, including the fiducial value of $\alpha_{z,P_0} = -0.758$. The fiducial value of $P_0$ is simply $P_0=1$ with this definition. We allow for further freedom in the gas pressure profile by also allowing the amplitude $\beta$ of the outer logarithmic slope $\tilde{\beta}$ to vary in the same manner as $P_0$ in Eq.~\ref{eq:P0def}:

\be
\label{eq:betadef}
\tilde{\beta}(M_{200},z) = \beta \left( \frac{M_{200}}{10^{14} \,\, M_{\odot}} \right)^{0.0393} \left( 1+z \right)^{0.415}  \,,
\ee

\noindent where the specific numbers are from the fitting function presented in \citet{BBPS2}, including the fiducial value of $\beta = 4.35$.

The $C_{\ell,\rmn{2h}}$ term describes the clustering of the sources responsible for the tSZ and lensing fields \citep{KK1999}. In the Limber approximation, which is highly accurate for the multipole range of interest here ($\ell > 100$), the two-halo term is \citep{HP2013}:

\be
\begin{aligned}
C^i_{\ell,\rmn{2h}} = & \int \dd z \frac{\dd V}{\dd z} P_\rmn{lin} \left(\frac{\ell + 1/2}{\chi(z)},z \right) \\
	& \int \dd M_1 \frac{\dd n}{\dd M_1} b(M_1,z) \tilde{y}_\ell (M_1,z) \\
	& \int dM_2 \frac{\dd n}{\dd M_2} b(M_2,z) \tilde{\phi}_{i,\ell} (M_2,z)
\end{aligned}
\ee

\noindent where $P_\rmn{lin}(k,z)$ is the linear matter power spectrum computed using CAMB\footnote{{\tt http://camb.info/}} and $b(M,z)$ is the linear halo bias from \citet{Tink2010}. Our integration limits are $0.005 < z < 10$ (or the upper redshift limit of the source galaxy distribution $p_s(z)$ in the galaxy lensing case) and $10^{5} \, \rmn{M}_{\odot}/h < M < 5 \times 10^{15} \, \rmn{M}_{\odot}/h$. We verify that all integrals converge with these limits.

\subsection{Simulations}
\label{sec:sims}

We simulated cosmological volumes ($L=165$ Mpc$/h$) using a modified version of the GADGET-2 smoothed particle hydrodynamics (SPH) code \citep{Gadget}.
This version of the GADGET-2 code includes sub-grid models for active galactic nuclei (AGN) feedback \citep{BBPSS}, radiative cooling, star formation, galactic winds, supernova feedback \citep{SpHr2003}, and cosmic ray physics \citep{2006MNRAS.367..113P,2007A&A...473...41E,2008A&A...481...33J}.
We used three variants of sub-grid models listed in order of increasing complexity:
\begin{itemize}
\item The non-radiative model with only gravitational heating (referred to as {\it shock heating}).
\item The model with radiative cooling, star formation, galactic winds, supernova feedback, and cosmic ray physics (referred to as {\it radiative cooling}).
\item The {\it radiative cooling} model with the addition of AGN feedback (referred to as {\it AGN feedback}).
\end{itemize}
Note that the {\it shock heating} model is not presented as a viable alternative to the other models, but as an extreme ICM model, since it has been shown to be significantly discrepant with group and cluster observations \citep[e.g.,][]{Puch2008,McCarthy2011,HBSBPS2013,HS2014}.
We ran a suite of simulations from ten unique initial conditions for each sub-grid model. The box sizes were 165 Mpc$/h$, with a resolution of 256$^3$ gas and DM particles, corresponding to a mass resolution of $M_\rmn{gas} = 3.2\times 10^{9} \, \rmn{M}_{\odot}/h$ and $M_\rmn{DM} = 1.54\times 10^{10} \, \rmn{M}_{\odot} /h$. The cosmological parameters used for these simulations were $\Omega_\rmn{M} = \Omega_\rmn{DM} + \Omega_\rmn{b} = 0.25$, $\Omega_\rmn{b} = 0.043$, $\Omega_\Lambda = 0.75$, $H_0=100\,h\,\rmn{km}\,\rmn{s}^{-1}\,\rmn{Mpc}^{-1}$, $h=0.72$, $n_\rmn{s} =0.96$ and $\sigma_8 = 0.8$. The {\it AGN feedback} model has subsequently been found to agree with local tSZ measurements of high-mass cluster pressure profiles \citep{PlnkP2013} and higher redshift X-ray measurements of massive cluster pressure profiles \citep{Mcdn2014}. Additionally, it is consistent with measurements of the stellar and gas content in low-redshift clusters \citep{BBPS3}, as well as the pressure profile inferred from X-ray stacking of low-redshift groups \citep{Sunetal2011}. Unless stated otherwise, we use the {\it AGN feedback} simulations as our fiducial sub-grid model.

\begin{figure*}
 \begin{minipage}[t]{0.50\hsize}
    \centering{\small $\yphi$:}
  \end{minipage}
  \begin{minipage}[t]{0.50\hsize}
    \centering{\small $\ygal$:}
  \end{minipage}
\begin{center}
  \hfill
  \resizebox{0.5\hsize}{!}{\includegraphics{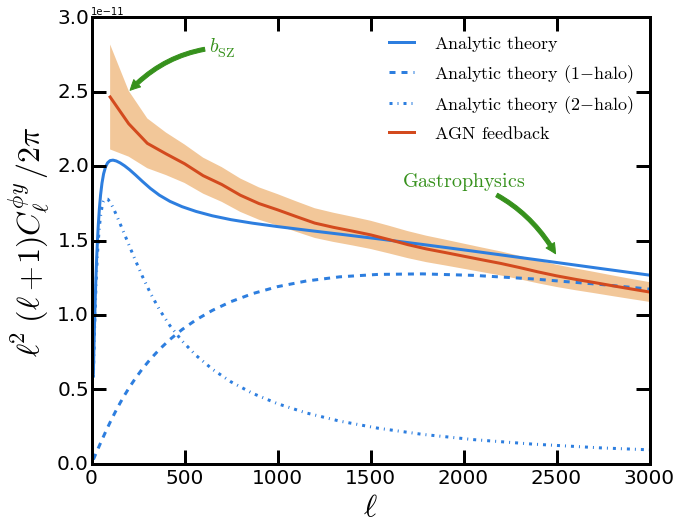}}%
  \resizebox{0.5\hsize}{!}{\includegraphics{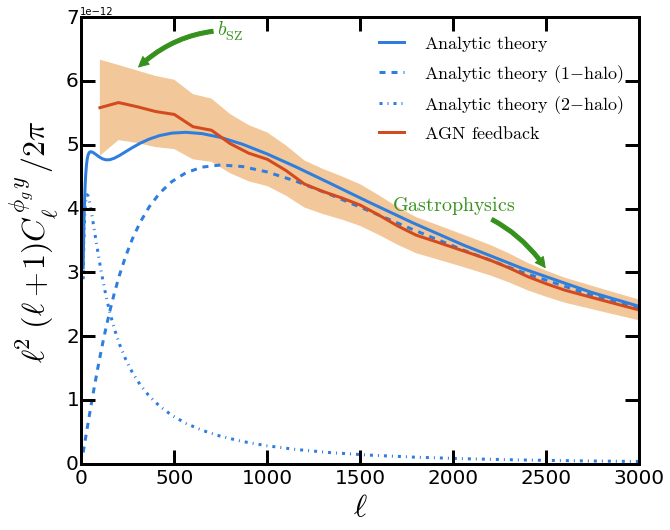}}\\
\end{center}
\caption{Comparison of the tSZ -- lensing cross-spectra from the analytic and simulation calculations. The left panel shows the tSZ -- CMB lensing cross-power spectrum $\yphi$, while the right panel shows the tSZ -- \emph{CFHTLenS} galaxy lensing cross-power spectrum $\ygal$. The one-halo, two-halo, and total contributions to the cross-spectrum (calculated analytically) are shown in dashed, dot-dashed, and solid blue lines, respectively. The shaded regions show the standard deviation about the average spectrum (red line) from ten different {\it AGN feedback} simulations. The cosmology, redshift limits, and pressure profile used for the analytic calculations match the simulation values, so only the total density profiles and mass functions differ between these calculations. The differences illustrated at high-$\ell$ in $\yphi$ result from baryonic effects on the density profiles, since the mass function only differs for the highest-mass halos at low redshift \citep{BBPS2}, which do not contribute significantly here (see Sec.~\ref{sec:mz}). At low-$\ell$, the differences in both spectra (seen more significantly in $\yphi$) likely arise from the presence of diffuse, unbound gas in the simulations, which is not captured in the analytic halo model calculations.}
\label{fig:thry}
\end{figure*}

We calculate the tSZ -- lensing cross-power spectra from the simulations as follows. Maps of the Compton-$y$ (Eq.~\ref{eq:y}) and the lensing convergence (Eq.~\ref{eq:sig}) signals are made at each redshift snapshot, from $z \approx 0.05 - 5$. We compute the cross-power spectrum for each redshift output from the $y$ and $\Sig$ maps and then average the cross-power spectra over the ten initial condition realizations. These average spectra are then summed over redshift\footnote{The simulations are written out at redshifts such that the step size equals the light crossing time of the simulation box length; thus, the total power spectrum is the sum of the differential power spectra.}. The advantages of this procedure are that it decreases the variance of the power spectrum and uses all the information within the simulation volume. Additionally, any correlations between different redshift slices are ignored, as effectively happens in nature, since the sum over redshift slices is taken {\it after} computing the power spectra.  

In each simulation, halo identification and characterization are required in order to calculate the cross-spectra as a function of halo mass, redshift, and cluster-centric radius. First, we find halos using a friends-of-friends algorithm \citep{Huch1982}. Then we iteratively compute each halo's center of mass and finally its spherical overdensity mass ($M_{\Delta}$) and radius ($R_{\Delta}$), as defined above. This procedure is performed at each redshift slice in the simulations. We use the resulting halo catalogs and their properties to deconstruct the tSZ -- lensing cross-spectra.

\section{Theory Results}
\label{sec:res}

Cross-correlations of the Compton-$y$ distortion and lensing fields are strong functions of cosmological parameters and halo properties \citep[][and Sec.~\ref{sec:thry}]{HS2014,vWHM2014}. Here, we fix the cosmological parameters to the values used in the simulations and exclusively quantify the dependence of the $\yphi$ and $\ygal$ cross-spectra signals on the properties of gas, stars, and DM in halos. We compare the cross-spectra from the {\it AGN feedback} simulations described in Sec.~\ref{sec:sims} to the analytic calculations described in Sec.~\ref{sec:thry} and interpret the resulting differences. We then use the full suite of simulations to deconstruct the contributions to the tSZ -- lensing cross-spectra as functions of ICM model, halo mass, redshift, and cluster-centric radius in order to better understand the physical origins of these cross-spectra.

\subsection{Comparison of the halo model to simulations}
\label{sec:comp}

To perform a like-for-like comparison between the halo model and the simulations, we implement the simulations' cosmological parameters (see Sec.~\ref{sec:sims}) and lower redshift cut at $z=0.05$ in the analytic calculations\footnote{The $z=0.05$ cut in the simulation calculations is necessary to reduce sample variance from rare, massive clusters in the derived power spectra \citep{Shawetal2009}.}. Furthermore, as described above, the analytic calculations use the pressure profile model derived from the {\it AGN feedback} simulations. Thus, any differences in the power spectra computed from these simulations and those computed from the halo model can only be due to quantities neglected in the halo model approximations, such as contributions from diffuse, unbound gas (e.g., in filaments) and changes to the halo density profile and halo mass function induced by baryonic effects. As a check on our calculations, we verify that the \emph{CFHTLenS} $\kappa$ auto-power spectrum computed from either the analytic calculations or the simulations agrees with that computed using the {\tt nicaea} code\footnote{http://www.cosmostat.org/nicaea.html} (and with one another). The agreement is nearly perfect in the linear regime and reasonably close in the non-linear regime, where baryonic effects could also be at work \citep[more detail will be discussed in][]{Batt2014prep}.

We first investigate the halo model results, before comparing them to the simulations. The one-halo and two-halo contributions to $\yphi$ and $\ygal$ are shown in Fig.~\ref{fig:thry}. The $\ygal$ cross-spectrum is computed for the \emph{CFHTLenS} source redshift distribution. For both cross-spectra, the term which dominates the signal is $\ell$-dependent . At low-$\ell$ (large angular scales), the two-halo term dominates. As $\ell$ increases, the cross-spectra transition to the one-halo term. The exact $\ell$ where this transition happens depends on the source redshift distribution $p_s (z_s)$. For $p_s (z_s)$ peaking at a higher redshift, the transition occurs at higher $\ell$ (smaller angular scales), as can be seen by comparing the transition points for $\yphi$ ($\ell \approx 500$) and $\ygal$ ($\ell \approx 150$). Fig.~\ref{fig:thry} illustrates that high signal-to-noise measurements of $\yphi$ and $\ygal$ over a wide multipole range will probe both the interior thermodynamic properties of halos (the one-halo term) and their global thermodynamic properties averaged over the cluster population (the two-halo term).

\begin{figure*}
 \begin{minipage}[t]{0.50\hsize}
    \centering{\small $\yphi$:}
  \end{minipage}
  \begin{minipage}[t]{0.50\hsize}
    \centering{\small $\ygal$:}
  \end{minipage}
\begin{center}
  \hfill
  \resizebox{0.5\hsize}{!}{\includegraphics{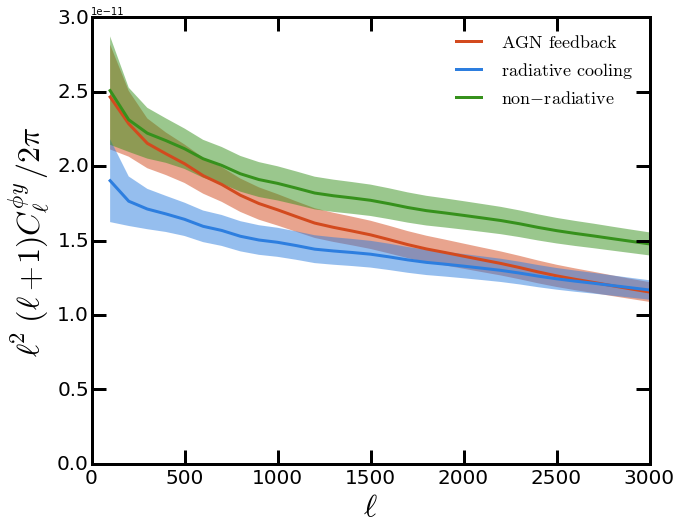}}%
  \resizebox{0.5\hsize}{!}{\includegraphics{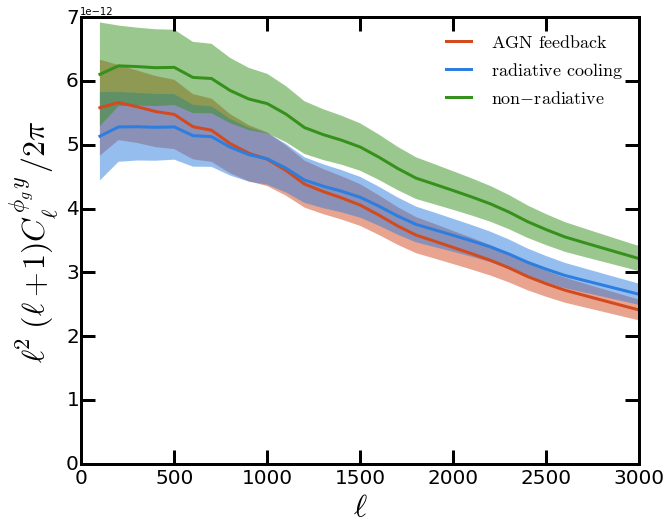}}\\
\end{center}
\caption{Dependence of tSZ -- lensing cross-spectra on the sub-grid gas model. The left panel shows $\yphi$ and the right panel $\ygal$ for {\it CFHTLenS}. Cross-spectra from the {\it shock heating} (labeled ``non-radiative''), {\it radiative cooling}, and {\it AGN feedback} simulations are shown by green, blue, and red lines, respectively. The shaded regions show the standard deviation about the average cross-spectra for the ten different simulation realizations of each model. The differences between the {\it shock heating} and {\it radiative cooling} simulations are the result of star formation removing halo gas and decreasing the total Compton-$y$ signal. At low-$\ell$ the cross-spectra from the {\it AGN feedback} simulations approach the {\it shock heating} simulations due to additional heating of the ICM. At high-$\ell$ the inner regions of the total mass and pressure profiles from the {\it AGN feedback} simulations are shallower than those found in the other models, causing a reduction in power.}
\label{fig:phys}
\end{figure*}

The $\ell$ range where the simulation and analytic calculations agree (within the simulation uncertainties from ten realizations) are $\ell \approx 1000 - 2500$ and $\ell \gtrsim 400$ for $\yphi$ and $\ygal$, respectively. At high-$\ell$, where the one-halo term dominates, $\yphi$ and $\ygal$ probe the shapes of the projected pressure and density profiles of the halos.  For $\ygal$, the analytic and simulation calculations agree very closely in this regime. For $\yphi$, we find that the analytic calculation predicts a higher cross-spectrum amplitude than the simulations. The analytic calculation uses the average pressure profile provided by the simulations \citep{BBPS2}, and thus these differences likely arise from baryonic effects on the density profile. The mass contributions in this regime are dominated by halos for which the simulations' mass function agrees well with \citet{Tink2008} \citep[][and Sec.~\ref{sec:mz}]{BBPS2}. The analytic calculation uses an NFW density profile \citep{NFW1997}, which contains a cuspy $r^{-1}$ density profile in the interior. This profile differs from the simulations, which have a flatter interior density profile due to baryonic feedback effects \citep[][will explore this in more detail]{Batt2014prep}. Due to the different density profile shapes, the analytic cross-spectrum will have more power than the simulation cross-spectrum on angular scales where the interior density profiles begin to be resolved (high-$\ell$). The results in Fig.~\ref{fig:thry} indicate that these baryonic effects on the interior density profile are more significant in higher-redshift, lower-mass halos, because the $\ygal$ analytic calculation matches the simulations well at high-$\ell$, while the $\yphi$ does not (the following subsections demonstrate that $\yphi$ is more sensitive to higher-redshift, lower-mass halos than $\ygal$). Although the total signal is a convolution of pressure and mass profiles, a high signal-to-noise measurement of $\yphi$ and $\ygal$ combined with a measurement of the pressure profile could provide constraints on the density profiles of the halos probed by $\yphi$ and $\ygal$ (for a fixed cosmological model, unless degeneracies with cosmological parameters can be broken).

On large angular scales (small $\ell$) the cross-spectra probe the large-scale bias between ICM thermal energy and the matter distribution ($b_\rmn{SZ}$). This bias is a sensitive tracer of energetic feedback (due to AGN, SNe, and more exotic sources) for the halos that are probed by $\yphi$ and $\ygal$, since feedback alters the global thermal properties of these halos. There will be degenerate effects between the many models for feedback and a natural trade-off between heating and depleting of the ionized gas in halos, which increase or decrease the cross-spectrum signal, respectively. However, a high signal-to-noise measurement of $\yphi$ and $\ygal$ could differentiate between such models (see Sec.~\ref{sec:fore}).

In addition to feedback effects, the low-$\ell$ cross-spectra are potentially sensitive to the presence of diffuse, unbound gas (``missing baryons''), which would manifest as an underestimate of the signal in the halo model calculations (which do not include such gas) compared to the simulations. Fig.~\ref{fig:thry} indicates a weak preference for such gas in the large-angle $\ygal$ cross-spectrum ($\ell \lesssim 400$), but a stronger preference in the $\yphi$ cross-spectrum ($\ell \lesssim 1000$). The diffuse gas signal is small but non-negligible, contributing $\approx 15$\% of the total signal at $\ell \approx 500$. This result is sensible in the context of the deconstructed cross-spectra presented below, which show that $\yphi$ is more sensitive to gas in lower-mass, higher-redshift halos at larger cluster-centric radii than $\ygal$. Accounting self-consistently for this diffuse gas when interpreting the measured $\yphi$ will shift the inferred cosmological parameters ($\sigma_8$ and $\Omega_\rmn{M}$) slightly downward from the values found in \citet{HS2014} (see Sec.~\ref{sec:obscomp}). However, degeneracies between the cosmological parameters and gas physics model currently do not allow for a robust detection of the diffuse gas signal in $\yphi$, as its presence cannot be straightforwardly separated from other sources contributing to the total observed signal. We revisit these points in Sec.~\ref{sec:obscomp}.

\subsection{Dependence on sub-grid gas models}
\label{sec:icm}

The shape and amplitude of the $\yphiname$ and $\ygalname$ cross-spectra are sensitive to the ICM modeling. Changes in the ICM model will mainly affect the Compton-$y$ contribution to $\yphi$ and $\ygal$. Although extreme cases of energetic feedback can significantly affect halo mass profiles (and thus $\phi_i$), such sub-grid models are not considered in this work. The sub-grid models affect the Compton-$y$ parameter through changes to the electron pressure profile \citep{BBPSS,BBPS2}. The processes of radiative cooling and star formation remove ionized gas from the ICM by converting it into stars, while feedback mechanisms slow this process and heat the surrounding gas.
In Fig.~\ref{fig:phys}, we show how the halo gas models affect the cross-spectra. The stark differences between cross-spectra from the {\it shock heating} and {\it radiative cooling} simulations are the result of star formation, which removes gas from halos and lowers the overall $y$-signal \citep[e.g.,][]{Nagai2007,BBPS1,Kay2012,LeBM2014}. The overall amplitude of these cross-spectra are a function of the gas fraction in halos, which sets the pressure profile normalization. The introduction of energetic feedback in the {\it AGN feedback} simulations affects the cross-spectra differently depending on the multipole considered. At low-$\ell$, the cross-spectra from the {\it AGN feedback} simulations approaches the {\it shock heating} spectra. Here, the additional heating from AGN in the {\it AGN feedback} simulations counteracts the loss of gas to star formation, affecting the global thermodynamics probed by the two-halo term. At high-$\ell$, the {\it AGN feedback} simulation spectra are similar to the {\it radiative cooling} spectra and decrease in amplitude to higher $\ell$. This additional reduction in power results from a shallower pressure profile in the cores of halos in the {\it AGN feedback} simulation compared to that found in the other simulations. The {\it AGN feedback} simulations expel gas or halt its initial infall onto halos which results in flatter interior pressure profiles. These effects likewise flatten the interior density profile as well \citep{Batt2014prep}.

\subsection{Mass and redshift dependences}
\label{sec:mz}

\begin{figure*}
 \begin{minipage}[t]{0.50\hsize}
    \centering{\small $\yphi$:}
  \end{minipage}
  \begin{minipage}[t]{0.50\hsize}
    \centering{\small $\ygal$:}
  \end{minipage}
\begin{center}
  \hfill
  \resizebox{0.5\hsize}{!}{\includegraphics{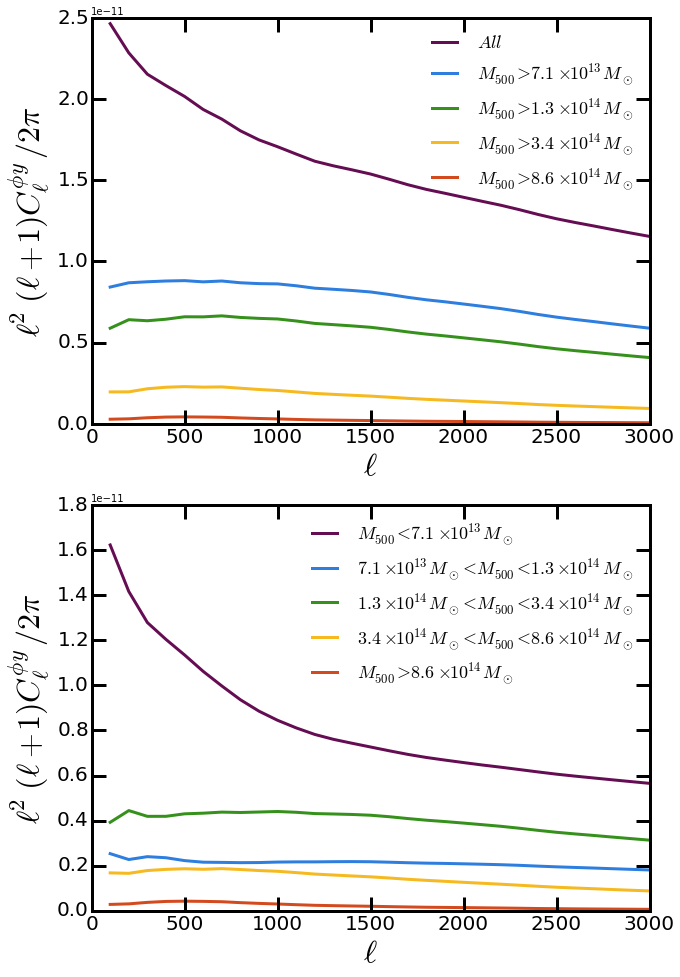}}%
  \resizebox{0.5\hsize}{!}{\includegraphics{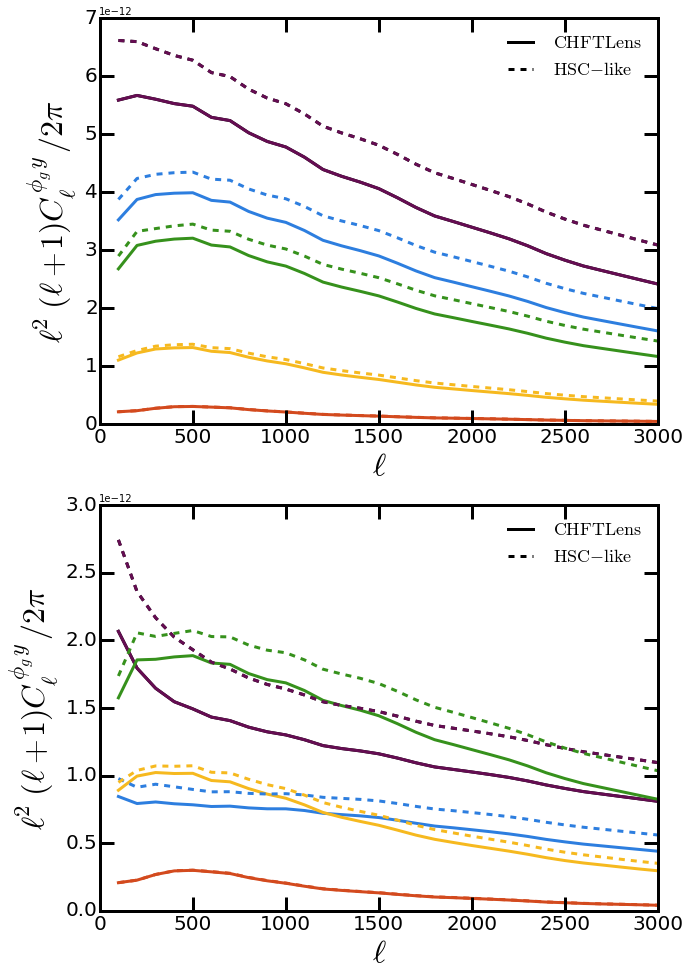}}\\
\end{center}
\caption{The tSZ -- lensing cross-spectra for various halo mass cuts in the {\it AGN feedback} simulations (the left panels show $\yphi$ and the right panels $\ygal$ for both {\it CFHTLenS} and {\it HSC}-like surveys). The top panels show the cross-spectra above a given halo mass threshold and the bottom panels show the signal within a given halo mass bin. Halos with $M_{500} < 7.1\times 10^{13} \, \rmn{M}_\sun$ contribute the most to $\yphi$. For $\ygal$, considering either {\it CFHTLenS} or {\it HSC}-like surveys (solid and dashed lines, respectively), halos with $1.3\times 10^{14} \, \rmn{M}_\sun < M_{500} < 3.4\times 10^{14} \, \rmn{M}_\sun$ contribute the most to the spectra. Thus, $\yphi$ is more sensitive to the gas in low-mass halos than $\ygal$, a result that can be traced to the different lensing kernels for these observables.}
\label{fig:mcut}
\end{figure*}

\begin{figure*}
 \begin{minipage}[t]{0.50\hsize}
    \centering{\small $\yphi$:}
  \end{minipage}
  \begin{minipage}[t]{0.50\hsize}
    \centering{\small $\ygal$:}
  \end{minipage}
\begin{center}
  \hfill
  \resizebox{0.5\hsize}{!}{\includegraphics{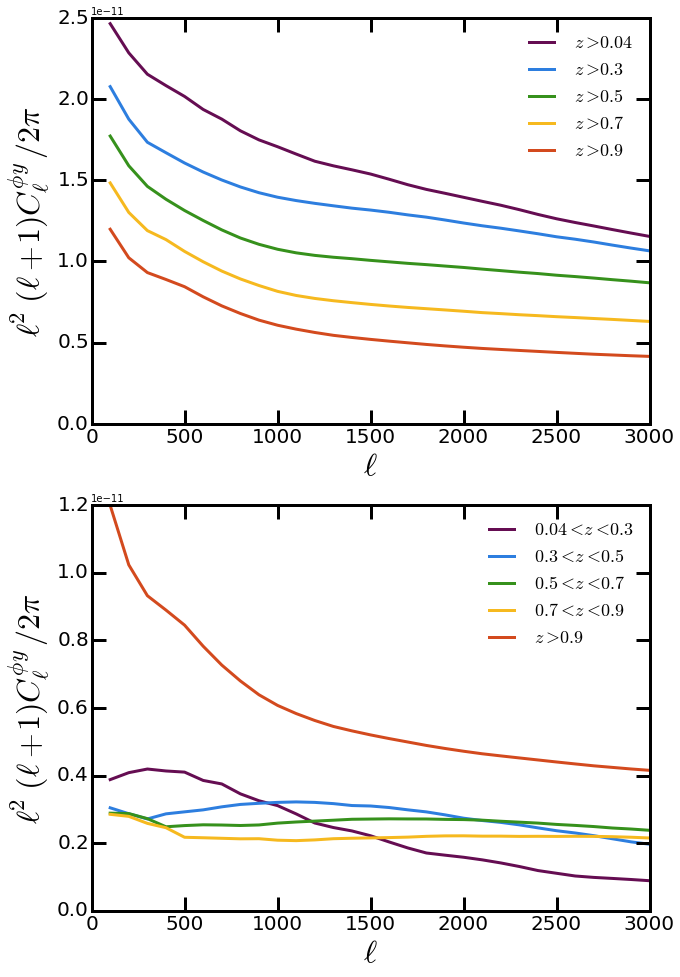}}%
  \resizebox{0.5\hsize}{!}{\includegraphics{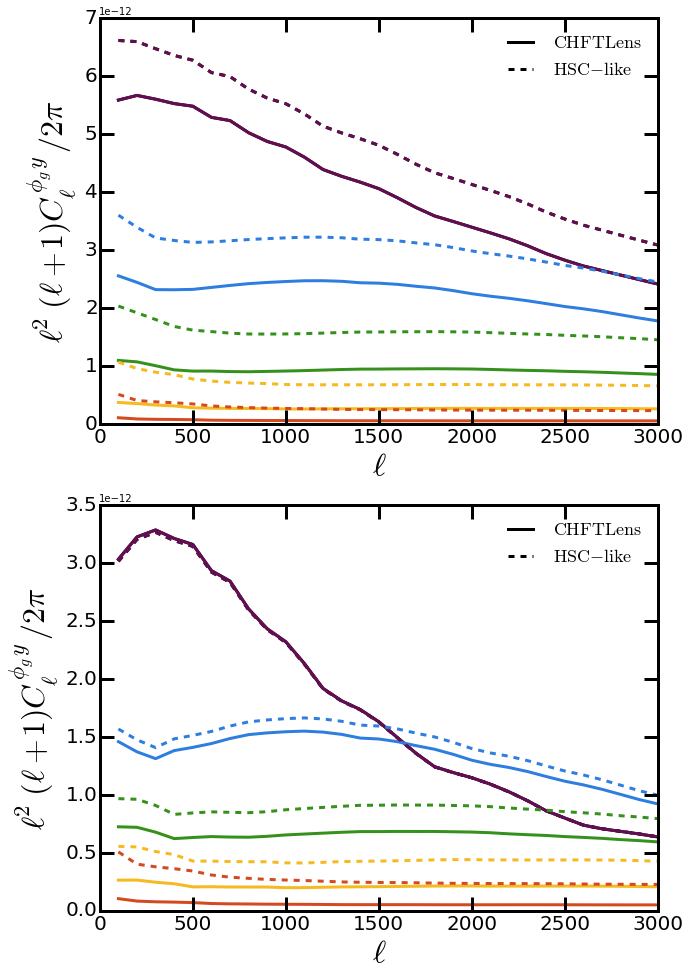}}\\
\end{center}
\caption{The tSZ -- lensing cross-spectra for various redshift cuts in the {\it AGN feedback} simulations (the left panels show $\yphi$ and the right panels $\ygal$ for both {\it CFHTLenS} and {\it HSC}-like surveys). The top panels show the cross-spectra below a given redshift and the bottom panels show the signal within a given redshift bin. Contributions from $z > 0.9$ dominate the $\yphi$ signal. For $\ygal$, considering either {\it CFHTLenS} or {\it HSC}-like surveys (solid and dashed lines, respectively), the redshift ranges $0.04 < z < 0.3$ at $\ell \lesssim 1500$ and $0.3 < z < 0.5$ at $\ell \gtrsim 1500$ contribute the most to the spectra. As expected due to the CMB lensing kernel, $\yphi$ probes higher redshifts than $\ygal$.}
\label{fig:zcut}
\end{figure*}

In this subsection, we deconstruct $\yphi$ and $\ygal$ in mass and redshift bins using the fiducial {\it AGN feedback} simulations. We consider both \emph{CFHTLenS} and \emph{HSC}-like source galaxy redshift distributions for $\ygal$. The mass and redshift deconstruction of $\yphi$ is also investigated in \citet{HS2014} in the halo model approximation, but not using simulations. We explore both cumulative and differential mass and redshift bins. We consider all gas particles (or radii) within $6 R_{500}$ when projecting the Compton-$y$ signal in the simulations. We use the full $\Sig$ maps. Our method is careful not to double-count the cluster mass in overlapping volumes of close-by cluster pairs.  Note that the halo mass cuts truncate the halo contribution at $6R_{500}$ (see Sec.~\ref{sec:rad} for details). This truncation removes some of the contributions to the two-halo term; thus, at low-$\ell$ where the two-halo term is important, the curves should be considered lower limits.

In Fig.~\ref{fig:mcut}, we show the cross-spectra $\yphi$ and $\ygal$ (left and right, respectively) broken down into cumulative (top panels) and differential (lower panels) mass bins. Fig.~\ref{fig:zcut} presents the analogous calculations for cumulative and differential redshift bins. The lensing kernels $\Wcmb$ and $\Wgal$ drive the differences in the mass and redshift dependences for $\yphi$ and $\ygal$. The Compton-$y$ signal is strongest for the most massive objects in the Universe, most of which do not form until late times ($z \lesssim 1$). The mass and redshift contributions to $\yphi$ and $\ygal$ arise from halos lying at the intersection of the relevant lensing kernel and the Compton-$y$ ``kernel'' driven by the formation of massive structures. Since the galaxy lensing kernel is restricted to low redshifts, larger halo masses contribute more to $\ygal$ than $\yphi$. Given the halo mass bins we choose, the largest contribution to $\ygal$ come from halos with $1.3 \times 10^{14} \, \rmn{M}_\sun < M_{500} < 3.4 \times 10^{14} \, \rmn{M}_\sun$ for $\ell \gtrsim 500$ for both {\it CFHTLenS} and {\it HSC}-like galaxy imaging surveys. In contrast, the largest contribution to $\yphi$ arises from halos with $M_{500} < 7.1\times 10^{13} \, \rmn{M}_\sun$ (given the mass bins we choose). This result is in agreement with that found in \citet{HS2014} (see their Fig.~5, convert mass definitions appropriately, and bin as in Fig.~\ref{fig:mcut} here).

The redshift cuts are easily understood in the context of the different lensing kernels. The \emph{CFHTLenS}, {\it HSC}-like, and CMB lensing kernels peak at increasingly higher redshifts, and thus the associated tSZ -- lensing cross-spectra probe gas at progressively higher redshifts. Given the redshift bins we choose, $\yphi$ is dominated by contributions from $z > 0.9$, while $\ygal$ (for either {\it CFHTLenS} or {\it HSC}-like) is mostly sourced by halos at $z < 0.3$ ($\ell \lesssim 1500$) or $0.3 < z < 0.5$ ($\ell \gtrsim 1500$). Note that the different source redshift distributions of different galaxy imaging surveys potentially allow for tomography of the tSZ -- lensing signal. For example, an {\it HSC}-like survey will have source galaxies to higher redshift than {\it CFHTLenS}, and thus its cross-spectrum is more sensitive to higher redshift and lower mass halos than {\it CFHTLenS}. Because of the different dependences of sub-grid physics models on mass and redshift, such tomographic measurements can potentially provide powerful mass- and redshift-dependent constraints on the ICM and feedback prescriptions \citep{Pratn2014}.

\subsection{Radial cuts}
\label{sec:rad}

\begin{figure}[t]
\plotone{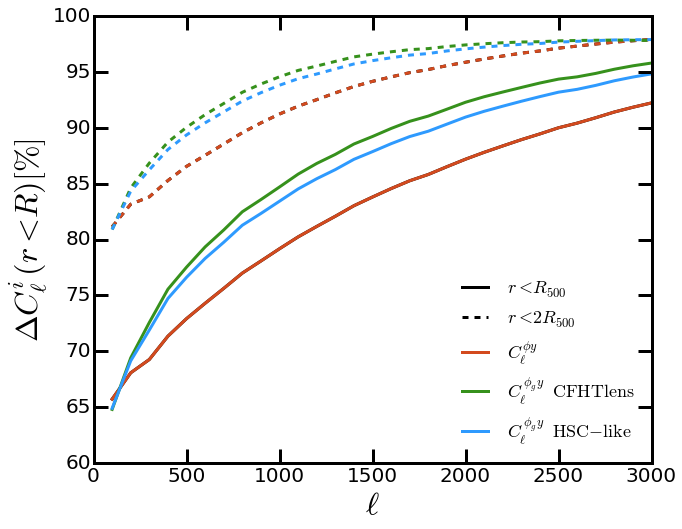}
\caption{The fractional contributions to $\yphi$ (red lines) and $\ygal$ (green and blue lines for {\it CFHTLens} and {\it HSC}-like surveys, respectively) for the radial truncations, $r < R_{500}$ (solid lines) and $r < 2R_{500}$ (dashed lines) on the {\it AGN feedback} simulations. Contributions beyond $r > R_{500}$ and $r > 2R_{500}$ are more important for $\yphi$ than $\ygal$. At low-$\ell$ the contributions from gas beyond $r > R_{500}$ and $r > 2R_{500}$ should be thought of as lower limits. At these angular scales the two-halo term dominates, and the outer regions of clusters contribute significantly to the cross-spectra. Where the one-halo term dominates the cross-spectra, the contribution from the outer region is not significant since the spectra are starting to resolve the halo interiors.}
\label{fig:radcut}
\end{figure}

We now investigate the regions of each halo contributing to the cross-spectra, to ascertain whether the core regions or the outskirts are responsible for the signals. We apply varying radial truncations to the simulated $y$-maps, using clusters with $M_{500} > 7.1 \times 10^{13} \, \rmn{M}_{\sun}$ at $0.05 < z <5$. We follow the procedure in \citet{BBPS2} to make real-space cuts and use a Gaussian taper when truncating at a given radius to avoid ringing in Fourier space. We place radial tapers at $r = R_{500}$, $2R_{500}$, and $6R_{500}$ in the $y$-maps, adopting $6R_{500}$ as the reference radial taper for the signal from the entire halo.

In Fig.~\ref{fig:radcut}, we show the fractional percentage contributions to $C_\ell^i$, defined as $\Delta C^i_{\ell} (r < R) \equiv 100 C^i_{\ell}(r < R)/ C^i_{\ell}(r < 6R_{500})$, where $C^i_{\ell}(r < 6R_{500})$ is the cross-spectrum from the $6R_{500}$ radial cut and $C^i_{\ell}(r < R)$ are cross-spectra from the other radial cuts. Note that since we cut the smaller halos with $M_{500} < 7.1 \times 10^{13} \, \rmn{M}_{\sun}$, we remove some of the two-halo term from the cross-spectra (similarly, contributions from diffuse gas are not included in the $C^i_{\ell}(r < 6R_{500})$ calculation). Thus, the percentages shown in Fig.~\ref{fig:radcut} for multipoles where the two-halo term dominates, $\ell \lesssim 500$ for $\yphi$ and $\ell \lesssim 150$ for $\ygal$, are upper limits to the contributions from within a given radius. For example, we find that gas at $r < R_{500}$ contributes $\lesssim 2/3$ of the power at the lowest multipoles. At $\ell \approx 3000$, this gas contributes $\approx 90$\% of the total power. Since the one-halo term dominates in this regime, the estimate should be accurate. We find that gas at $r > 2R_{500}$ contributes $\gtrsim 15$\% at low $\ell$ and $\approx 5$\% at high $\ell$. The contributions at large radii, $r > R_{500}$  and $r > 2R_{500}$, are greater for $\yphi$ then for $\ygal$, a result that can be traced to the different lensing kernels as in the previous subsection. At high $\ell$, we show that the cross-spectra are starting to resolve the halo centers and gas inside $R_{500}$ contributes an overwhelming majority of the power to the cross-spectra.

\section{Comparison to observations}
\label{sec:obscomp}

\begin{figure}[t]
\plotone{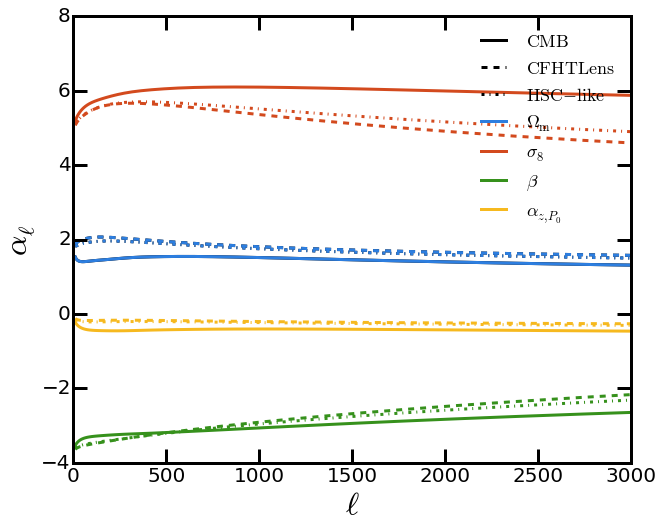}
\caption{Power-law scaling $\alpha^q_{\ell}$ of $\sigma_8$, $\Omega_\rmn{M}$, $\beta$, and $\alpha_{z,P_0}$ for the cross-spectra $\yphi$ and $\ygal$ as a function of $\ell$ (see Eq.~\ref{eq:asig}). The cross-spectra scale linearly with the normalized amplitude $P_0$ by definition, so it is not plotted for clarity. The power-law scaling $\alpha^q_{\ell}$ is roughly constant for most parameters across the $\ell$ range shown, but we use the full $\ell$-dependent function for each parameter in this work.}
\label{fig:alpha}
\end{figure}

The initial $\approx 6 \sigma$ measurements of $\yphi$ \citep{HS2014} and $\ygal$ \citep{vWHM2014} fit the data individually using different models \citep[see][respectively]{HS2014,Ma2014}. Here, we re-interpret the measurements in the context of the {\it AGN feedback} model discussed in Section~\ref{sec:meth}, using both analytic halo model calculations (which match the procedure used in \citet{HS2014}) and simulations. We choose $\sigma_8 = 0.817$ and $\Omega_\rmn{M} = 0.282$ as the fiducial cosmological parameter values (these are the WMAP9+eCMB+BAO+$H_0$ maximum-likelihood parameters \citep{Hinshawetal2013}, which we refer to as the WMAP9 cosmology for brevity). The fiducial gas physics parameters are $P_0 = 1$, $\beta = 4.35$, and $\alpha_{z,P_0}  = -0.758$ as described in Section~\ref{sec:thry}, matching the {\it AGN feedback} model. The fiducial parameter set is denoted as $p^q_0$ where $q$ labels each parameter, i.e., $q \in \left\{ \sigma_8, \Omega_\rmn{M}, P_0, \beta, \alpha_{z,P_0} \right\}$. We then use the analytic halo model calculations to compute the dependence of the tSZ -- lensing cross-spectra on each parameter. Thus, we use the fiducial analytic cross-spectra, $C^i_{\ell}(p^q_0)$, and compute new spectra by perturbing only one parameter in a given calculation. At each multipole $\ell$, we compare the relative amplitudes of the spectra

\be
\label{eq:asig}
C^i_\ell (p^q) = C^i_\ell(p^q_0) \left(\frac{p^q}{p^q_0}\right)^{\alpha^q_\ell },
\ee

\noindent where $p^q$ is the perturbed parameter. Here, we assume that the cross-spectra scale as a power-law function of the perturbed parameter at each $\ell$, with a power-law index $\alpha^q_{\ell}$. In Figure~\ref{fig:alpha}, we show the values for $\alpha^q_\ell$ for each parameter in the model. Changes in $P_0$ scale linearly into changes in $\yphi$ and $\ygal$ (c.f., Eq.~\ref{eq:P0def}) and thus are not shown for clarity. The most sensitive parameter, as expected from previous tSZ studies \citep[e.g.,][]{KS2002,HP2013,HS2014}, is $\sigma_8$, with the cross-spectra scaling roughly as $\sigma_8^{5-6}$ over the $\ell$ range considered.

\begin{figure*}
 \begin{minipage}[t]{0.50\hsize}
    \centering{\small $\yphi$:}
  \end{minipage}
  \begin{minipage}[t]{0.50\hsize}
    \centering{\small $\ygal$:}
  \end{minipage}
\begin{center}
  \hfill
  \resizebox{0.5\hsize}{!}{\includegraphics{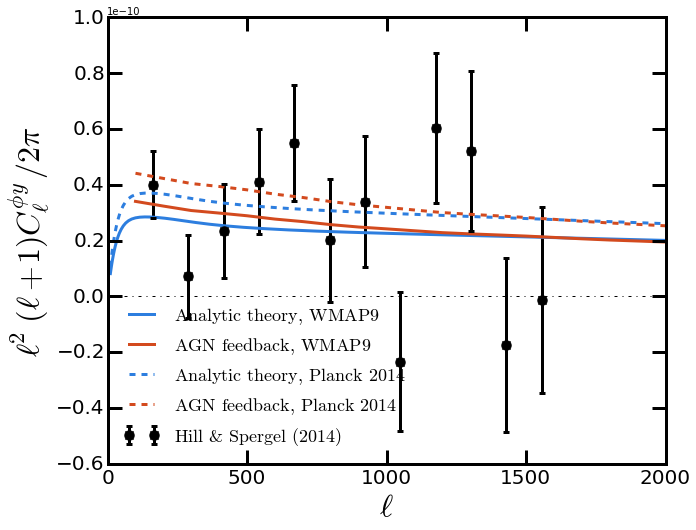}}%
  \resizebox{0.5\hsize}{!}{\includegraphics{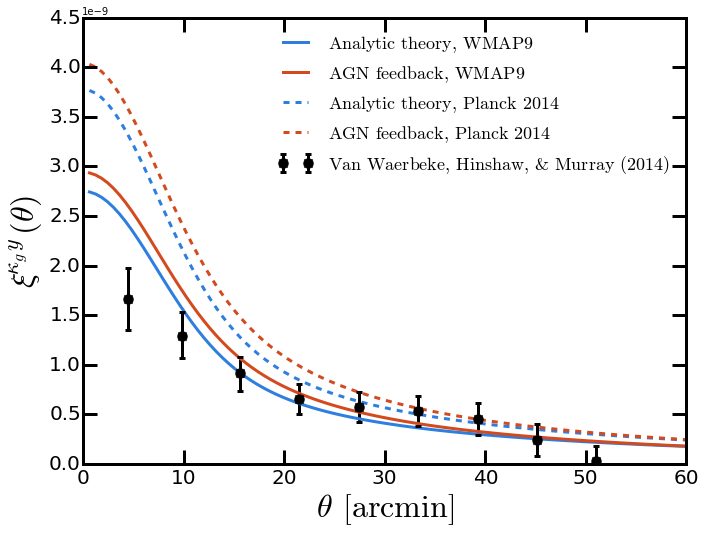}}\\
\end{center}
\caption{Comparison of the cross-spectra from the {\it AGN feedback} simulations and analytic halo model calculations to the observational results from \citet{HS2014} and \citet{vWHM2014}. In the right panel, we convert the theoretical results to the real-space cross-correlation function $\xi^{\kappa_g y}(\theta)$ from \citet{vWHM2014}. Both measurements prefer a lower amplitude than that predicted by the \emph{Planck} cosmological parameters. Note that the multipole-space $\yphi$ data points in the left panel are nearly uncorrelated from bin to bin, while the real-space $\xi^{\kappa_g y}$ data points in the right panel are strongly correlated.}
\label{fig:obcomp}
\end{figure*}

Using the dependence of the cross-spectra on each parameter, we investigate fits to the $\yphiname$ and $\ygalname$ measurements. In Fig.~\ref{fig:obcomp}, we compare the simulation and analytic theory results from the previous section to the data. The measurement of $\ygal$ is made in terms of a real-space cross-correlation function $\xi^{\kappa_g y}(\theta)$ of Compton-$y$ and {\it CFHTLenS} $\kappa_g$, and we thus Legendre transform the $\ygal$ theory and convert $\phi_g$ to $\kappa_g$ appropriately. In the Legendre transformation, we also account for the smoothing of the $y$ and $\kappa_g$ maps used in the measurement \citep{vWHM2014}\footnote{Note that the FWHM of the $\kappa_g$ map is 9.9 arcmin (L. van Waerbeke, priv. comm.).}. We extend the simulation curve to the lowest multipoles needed for the Legendre transformation by assuming a smooth interpolation based on the analytic results. In both panels of Fig.~\ref{fig:obcomp}, the small differences between the simulation and analytic calculations result from the effects described in Sec.~\ref{sec:comp}, specifically the signal from diffuse, unbound gas and the flattening of the inner density profile due to baryonic feedback. Note that these effects are convolved in the real-space cross-correlation shown in the right panel of Fig.~\ref{fig:obcomp}.

More important, however, is the role of cosmological parameter variations. For this exercise, we leave the gas physics model fixed to the {\it AGN feedback} prescription, and consider WMAP9 or \emph{Planck} values for $\sigma_8$ and $\Omega_\rmn{M}$. The \emph{Planck} values are $\sigma_8 = 0.831$ and $\Omega_\rmn{M} = 0.316$ \citep{PlanckParams}. For the $\yphiname$ results, we compute simple $\chi^2$ values for the simulation curves with respect to the measured $\yphi$ data. The simulation results include effects neglected in \citet{HS2014}, such as the presence of diffuse, unbound gas at large angular scales (low $\ell$). We find $\chi^2 = 14.2$ and $\chi^2 = 16.9$ for the WMAP9 and \emph{Planck} cosmological parameters, respectively, with 12 degrees of freedom in either case. Thus, in the context of the {\it AGN feedback} pressure profile model, the $\yphi$ data moderately prefer the WMAP9 parameters to those from \emph{Planck}. This result matches the qualitative conclusions of \citet{HS2014}, although the preference for WMAP9 over \emph{Planck} is stronger here because of the higher $\yphi$ predicted by the simulations for a given set of cosmological parameters. To compare further with the results of \citet{HS2014}, we fit the best-determined degenerate combination of $\sigma_8$ and $\Omega_\rmn{M}$. The best-fit result is $\sigma_8 (\Omega_\rmn{M} / 0.282)^{0.26} = 0.814$ with $\chi^2 = 14.2$, nearly identical to the WMAP9 value, with an error bar matching the result from \citet{HS2014} of $\sigma_8 (\Omega_\rmn{M} / 0.282)^{0.26} = 0.824 \pm 0.029$. Thus, as expected due to the inclusion of signal missing in the halo model calculations of \citet{HS2014}, the best-fit amplitude has decreased slightly, although well within the statistical error bar.

We perform similar exercises for the $\xi^{\kappa_g y}(\theta)$ measurements of \citet{vWHM2014}, though only at a qualitative level, as $\chi^2$ values cannot be robustly computed without using the full covariance matrix for this observable (i.e., the points are significantly correlated), which is not publicly available. Fig.~\ref{fig:obcomp} compares the {\it AGN feedback} analytic and simulation calculations for both WMAP9 and \emph{Planck} parameter values to the measurements. The \emph{Planck} calculations are clearly much higher than the data, especially at small angular scales. The tension is somewhat relieved by using WMAP9 parameters. The small-scale data points can be better fit with lower values of $\sigma_8$ and $\Omega_\rmn{M}$ (e.g., $\sigma_8 = 0.8$ and $\Omega_\rmn{M} = 0.25$, the values used in the \citet{BBPSS} simulations), a result that agrees with direct cluster count measurements \citep[e.g.,][]{Hass2013,Planckcounts}, tSZ power spectrum measurements \citep[e.g.,][]{Siev2013,George2014,PlnkY2013}, and measurements of higher-order tSZ statistics \citep{Wilson2012,Crawford2014,Hill2014}. However, the better fit at small scales comes at the cost of a slightly worse fit to the large-scale data points. The large scales can possibly be further remedied by modifying the pressure profile model or including additional diffuse, unbound gas --- but clearly these possibilities are degenerate with changes in the cosmological parameters.

At large angular scales in $\xi^{\kappa_g y}(\theta)$ (corresponding to low-$\ell$ in $\ygal$), the halo model and simulation calculations agree well, with less evidence for diffuse, unbound gas (``missing baryons'') than in the $\yphi$ calculations --- see Fig.~\ref{fig:thry}. Thus, for a WMAP9 or \emph{Planck} cosmology, the large angular scales in $\xi^{\kappa_g y}(\theta)$ do not require additional signal (in fact the \emph{Planck} prediction is already too high); for different cosmological parameters, this conclusion will vary, thus reflecting the degeneracy between changes in the gas physics model and cosmology that affects nearly all tSZ measurements, including cluster counts \citep[e.g.,][]{Hass2013,Planckcounts} and indirect statistics \citep[e.g.,][]{HP2013,McCarthy2014,HS2014}. A robust detection of the missing baryons (diffuse, unbound gas beyond halos) in an observed tSZ -- lensing cross-correlation would require a demonstration that the data can only be well fit when including the excess power at low-$\ell$ seen in the simulations over the halo model prediction (see Fig.~\ref{fig:thry}), and that changes to the gas pressure profile model or cosmological parameters cannot be made instead to improve the fit. Clearly the current tSZ -- lensing cross-correlation measurements are far from this regime, given the error bars and significant outstanding uncertainty on the gas pressure profile model.

\section{Future observational constraints}
\label{sec:fore}

In this section, we forecast the ability to simultaneously constrain cosmological and astrophysical parameters by combining $\yphiname$ and $\ygalname$ measurements. We use the current measurements of $\yphi$ and $\ygal$ as a baseline, and anticipate the expected improvements in signal-to-noise over these measurements from ongoing and future experiments. We use the Fisher matrix formalism \citep[e.g.,][]{Fisher1935,Knox1995,Jung1996} to forecast the expected constraints on these parameters. As with all Fisher analyses, we assume gaussian errors. We also assume that $\yphi$ and $\ygal$ are well described by the halo model described in Sec.~\ref{sec:thry} and that the parameters used in the modeling (both cosmological and astrophysical) are reasonably close to the {\it real} values.  The Fisher matrix $F_{jk} $ is calculated 

\be
F_{jk} = \frac{\dd C^i_{\ell}}{\dd p_j}  (M^{-1})_{\ell \ell'} \frac{\dd C^i_{\ell'}}{\dd p_k} 
\ee

\noindent where $(M^{-1})_{\ell \ell'} $ is the inverse covariance matrix and $p_j$ is $j^\rmn{th}$ parameter that we are forecasting. 
We calculate $M_{\ell \ell'}$ using pure statistical errors bars for the cross-spectra,

\be
\left(\Delta C^{1,2}_\ell \right)^2 = \frac{1}{f_\rmn{sky} (2\ell+1)\Delta \ell} (C_\ell^1C_\ell^2 + C_\ell^{1,2}),
\ee

    \noindent where $f_\rmn{sky}$ is the observed fraction of the sky, $\Delta \ell$ is the bandpower width, and $C_\ell^1$, $C_\ell^2$, and $C_\ell^{1,2}$ are the {\it observed} auto and cross-spectra (including the noise biases). For $C_\ell^{yy}$ we use the observed spectrum from \citet{HS2014} (which includes the significant non-tSZ noise bias) and we estimate a signal-to-noise improvement of $\approx \sqrt(5/2)$ in the final data release from \emph{Planck}. In this analysis, the fiducial $y-$map is denoted by $y^{1\rmn{st}}$ and the future, improved $y-$map is denoted by $y^{2\rmn{nd}}$. Forecasting the signal-to-noise of future $y-$maps with improved component separation techniques is beyond the scope of this paper (see \citet{HP2013} for an example).

We use the theoretical predictions of $\yphi$ and $\ygal$ computed in Section~\ref{sec:meth} for $C_\ell^{1,2}$, which assume that the cross-spectra contain pure tSZ -- lensing signal. We use the measured CMB lensing power spectrum from \emph{Planck} \citep{PLNKLens} for our initial estimate of $C_\ell^{\phi\phi}$ (including the noise bias). We estimate future CMB $C_\ell^{\phi\phi}$ plus noise using the minimum-variance estimator from \citet{HO2002} for Stage 2 CMB experiments (e.g., ACTpol and SPTpol) and Stage 3 CMB experiments (e.g., AdvACT and SPT3G). We estimate the observed galaxy lensing convergence auto-power spectrum $C_\ell^{\kappa \kappa, \rmn{obs}}$ as,

\begin{figure*}
 \begin{minipage}[t]{0.50\hsize}
    \centering{\small Current}
  \end{minipage}
  \begin{minipage}[t]{0.50\hsize}
    \centering{\small Future}
  \end{minipage}
\begin{center}
  \hfill
  \resizebox{0.5\hsize}{!}{\includegraphics{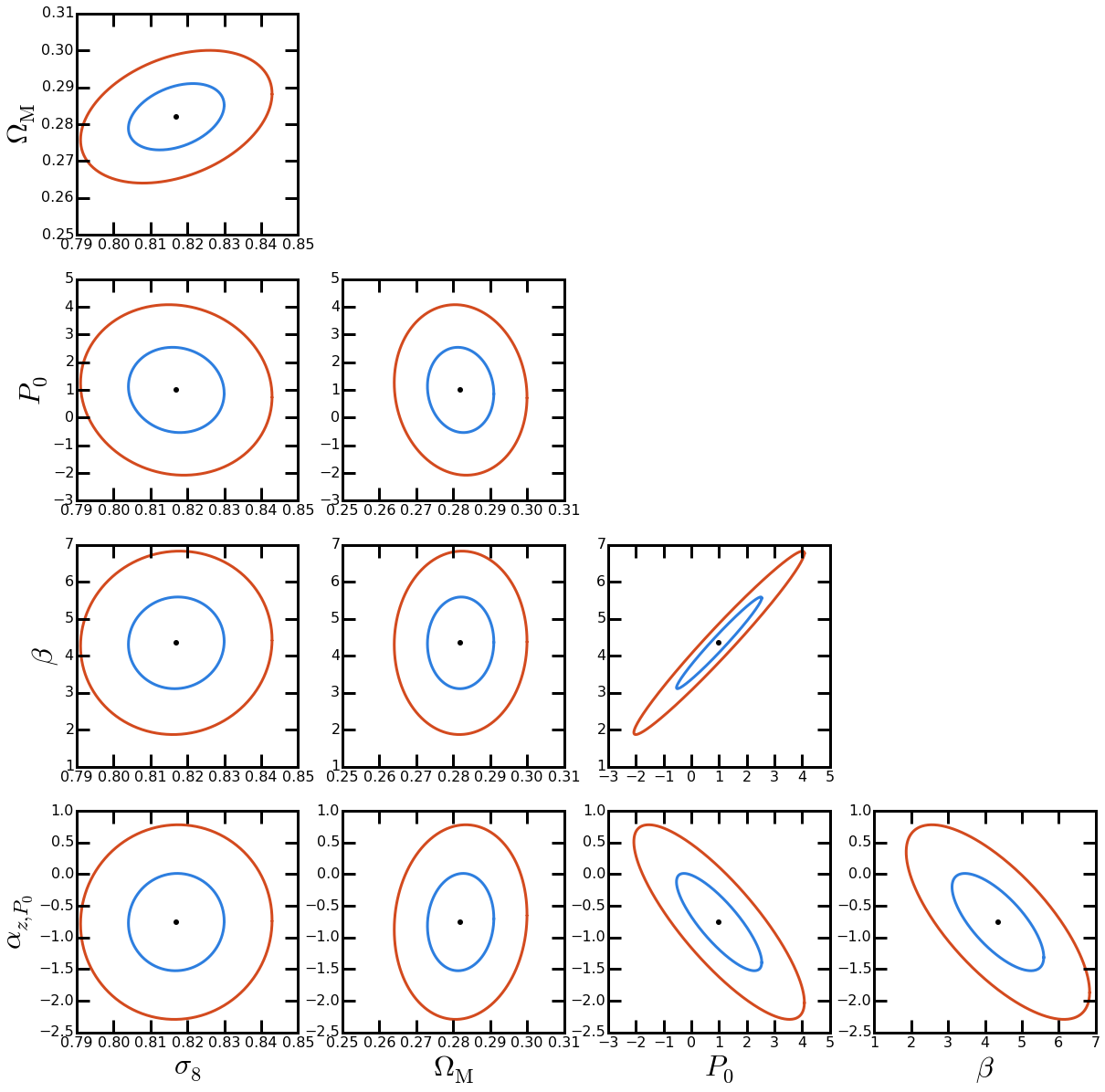}}%
  \resizebox{0.5\hsize}{!}{\includegraphics{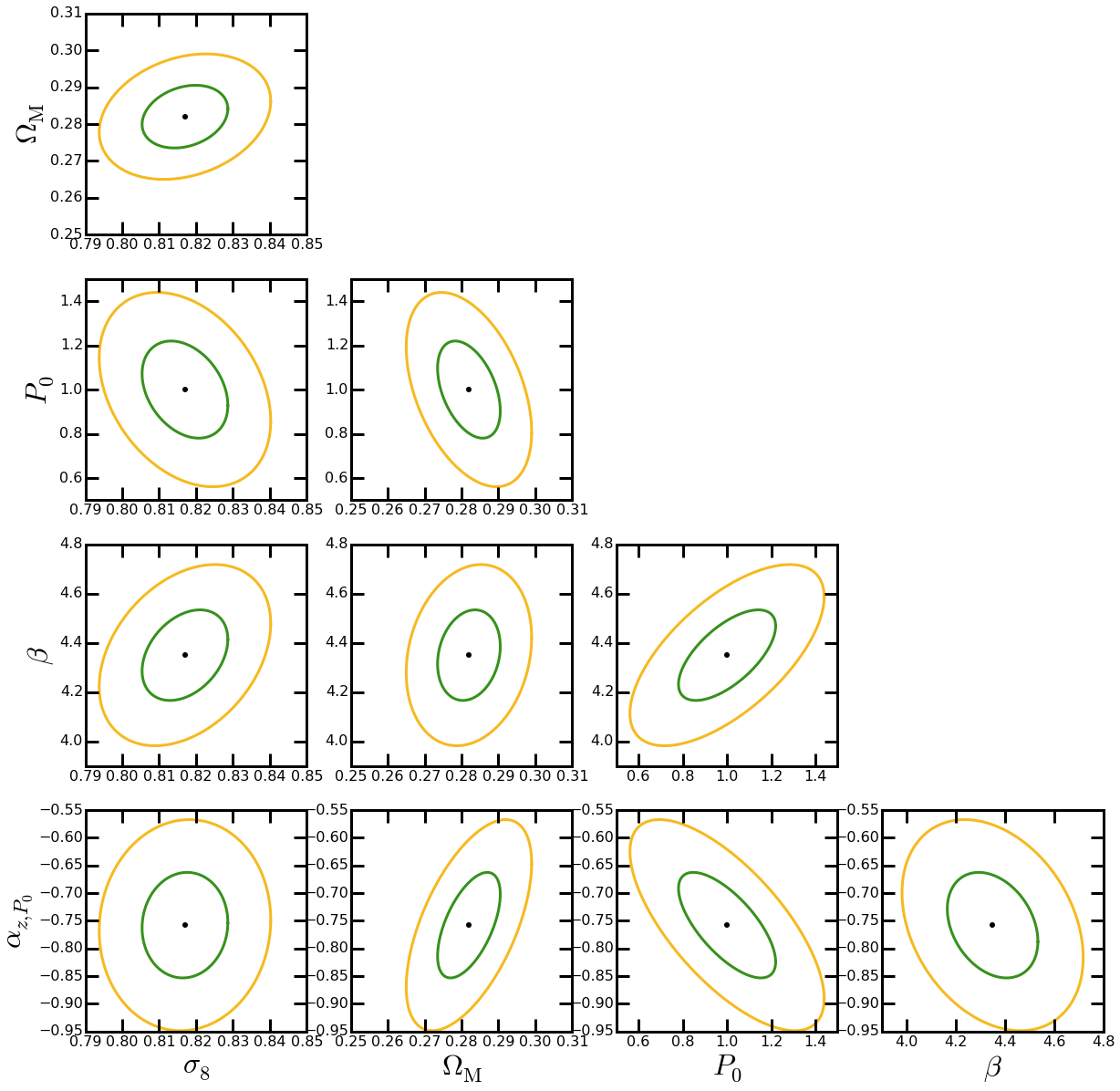}}\\
\end{center}
\caption{Fisher forecast for current and future constraints on cosmological and ICM parameters from the combination of $\ygal$ and $\yphi$ measurements. In both panels, constraints from the \emph{Planck} primary CMB are included to break parameter degeneracies. The ellipses denote $1\sigma$ and $2\sigma$ confidence levels.}
\label{fig:fish}
\end{figure*}

\be
C_\ell^{\kappa\kappa, \rmn{obs}} = C_\ell^{\kappa\kappa} + \frac{\sigma_\gamma^2 }{n_s},
\ee

\noindent where $\sigma_\gamma^2 / n_s$ is the shape noise term, which results from the finite number of source galaxies that are averaged over. The values for $\sigma_\gamma$, the intrinsic ellipticity dispersion per component, and $n_s$, the number of source galaxies per square arcminute, will depend on the survey. In Table \ref{tab:pars} we summarize the values used for each survey. For galaxy lensing we consider \emph{CFHTLenS}, Stage 3 ground-based surveys (e.g., HSC and DES), a Stage 4 ground-based survey (LSST), and a Stage 4 satellite survey (\emph{Euclid}) .

\begin{table*}
\caption[Specifications for tSZ -- lensing cross-correlation experiments considered in the Fisher analysis. The second line labels the lensing experiment under consideration, while the bottom line labels the $y$-map considered.]{Specifications for tSZ -- lensing cross-correlation experiments considered in the Fisher analysis.}
\begin{center}
\begin{tabular}{c|cccc|ccc} 
\hline \hline
&&Galaxy&&&CMB& \\
\hline
Experiments &\emph{CFHTLenS}&Stage 3& LSST & \emph{Euclid} & \emph{Planck} & Stage 3\\
\hline
$f_\rmn{sky}$ &0.005&0.048&0.25&0.2&0.25177& 0.25\\
$\sigma_\gamma^2$ &0.28&0.28&0.28&0.22&-&- \\
$n_s \, [{\rm arcmin}^{-2}]$ &7.6&15&40&35&-&- \\
$y-$map &$y^{1\rmn{st}}$&$y^{2\rmn{nd}}$&$y^{2\rmn{nd}}$&$y^{2\rmn{nd}}$&$y^{1\rmn{st}}$& $y^{2\rmn{nd}}$\\
\hline \hline
\end{tabular}
\end{center}
\label{tab:pars}
\end{table*}

To combine the experiments, we sum the different $F_{jk}$, which assumes that the measurements of $\yphi$ and $\ygal$ are uncorrelated.
This assumption is valid as long as we do not use surveys with overlapping sky coverage\footnote{We neglect small correlations due to common long-wavelength modes.}. Any overlap will result in the measurements using the same objects in the $y$-map and/or the density field, and thus the measurements will no longer be uncorrelated. In the cases where the surveys would overlap, we enforce the constraint that each survey has a unique survey area, so that we do not double-count the information. For related reasons, we also do not include information from the auto-power spectra of the Compton-$y$ or lensing measurements, although these clearly possess constraining power. We leave a full analysis of the joint covariances of the tSZ auto-, lensing auto-, and tSZ -- lensing cross-power spectra for future work.

We forecast constraints on five parameters, two cosmological and three astrophysical, as listed in Section~\ref{sec:obscomp}. The cosmological parameters we consider are $\sigma_8$ and $\Omega_\rmn{M}$, which both strongly influence the number of halos as a function of mass and redshift. For the astrophysical parameters, we reduce the large range of uncertainties in modeling the halo gas to three pressure profile parameters. In principle, the halo density profiles will also be altered due to changes in feedback and star formation modeling, but these effects will be sub-dominant to changes in the pressure profiles. From Eqs.~\ref{eq:P0def} and~\ref{eq:betadef}, we vary $P_0$, $\beta$, and $\alpha_{z,P_0}$. The parameter $P_0$ governs the total amount of thermal energy in a halo. Removal of gas into stars via star formation will decrease $P_0$, while heating of the gas via feedback will increase it. The $\beta$ parameter controls the outer logarithmic slope of the profile, which is sensitive to the amount of feedback in halos. Finally, the parameter $\alpha_{z,P_0}$ controls the redshift evolution of the total amount of thermal energy in halos and is sensitive to departures from the standard redshift evolution predicted by self-similar collapse \citep{Kaiser1986}. In the Fisher analysis, we use the complete $\ell$-dependent results for the power-law scalings of the cross-spectra with respect to each parameter, $\alpha^j_\ell$, as computed in Section~\ref{sec:obscomp}.

In Figure~\ref{fig:fish}, we show the estimated parameter constraints for two combinations of $\yphi$ and $\ygal$ measurements. The first (left panel) represents the {\it current} measurements: a combination of $\ygal$ from \emph{CFHTLenS} \citep{vWHM2014}, $\ygal$ from a Stage 3 galaxy lensing survey, and $\yphi$ from \emph{Planck} \citep{HS2014}. To break parameter degeneracies, we include the primary CMB constraints on $\sigma_8$ and $\Omega_\rmn{M}$ from \emph{Planck} \citep{PlanckParams}. Note that the constraints on $\sigma_8$ and $\Omega_\rmn{M}$ are completely driven by the \emph{Planck} primary constraints. If we had instead placed strong priors on the gas physics parameters, the tSZ -- lensing data could yield improvements in the cosmological constraints. However, our focus here is on using the tSZ -- lensing measurements to learn about the ICM, and thus we place no priors on the gas physics parameters. In this framework, current data are mostly useful for constraining the gas physics model.

In the right panel of Figure~\ref{fig:fish}, we show the constraints with the combination of the \emph{Euclid} satellite, LSST, and Stage 3 CMB experiments (for estimated AdvACT sky coverage). These surveys will cover approximately half the sky or more, but we assume that each of them uniquely covers only a fourth of the sky. Therefore, each measurement is independent and their Fisher matrices can be summed without considering the covariances between them. These forecasts also include the primary CMB constraints on $\sigma_8$ and $\Omega_\rmn{M}$ from \emph{Planck} \citep{PlanckParams}. The constraints on the astrophysical parameters are much tighter than those forecast for current experiments, and the tSZ -- lensing cross-correlation data now tighten the constraints on $\sigma_8$ and $\Omega_\rmn{M}$ slightly as well. As noted above, if we had placed priors on the gas physics parameters, the tSZ -- lensing measurements would provide significant additional constraining power on the cosmological parameters beyond the \emph{Planck} primary CMB data. But with no such priors in place, degeneracies between the gas physics and cosmological parameters result in the tSZ -- lensing data mostly improving constraints on the gas physics parameters --- to a very promising level of precision.

We summarize in Table \ref{tab:res} the fully marginalized constraints on the cosmological and astrophysical parameters. Although ongoing and near-future measurements of the cross-correlations yield fairly weak constraints on the astrophysical parameters, the forecast for future experiments is much more promising. We find marginalized fractional errors of $\approx 22$\%, $\approx 4$\%, and $\approx 13$\% on $P_0$, $\beta$, and $\alpha_{z,P_0}$, respectively (recall that the fiducial values are $P_0 = 1$, $\beta = 4.35$, and $\alpha_{z,P_0} = -0.758$) . With these potential constraints, it will be possible to start to distinguish between sub-grid ICM models for star formation and feedback.

\begin{table}
\caption[Marginalized errors on parameters.]{Marginalized errors on parameters.}
\begin{center}
\begin{tabular}{c|c|c} 
\hline \hline
Parameters & Current & Future \\
\hline
$\Delta \sigma_8$ &0.013 (1.6\%)& 0.012 (1.4\%)\\
$\Delta \Omega_\rmn{M}$ &0.0090 (2.8\%)& 0.0085 (2.7\%)\\
$\Delta P_0$ &1.9 (190\%)&0.22 (22\%)\\
$\Delta \beta$ &1.5  (34\%) &0.18 (4.1\%)\\
$\Delta \alpha_{z,P_0}$ &1.1 (150\%)&0.095 (13\%)\\
\hline
\end{tabular}
\end{center}
\label{tab:res}
\end{table}

\section{Conclusions}
\label{sec:conc}
How hot, ionized gas traces the underlying mass in the Universe is an important cosmological and astrophysical question. Weak lensing observations robustly trace the matter distribution, while tSZ observations track the thermal pressure of hot, ionized gas. Naturally, the cross-correlation of these quantities probes the interplay between the mass and ionized gas. Recently, the cross-correlation of the these quantities was measured at $\approx 6\sigma$ independently by \citet{HS2014} and \citet{vWHM2014}, by cross-correlating independently constructed Compton-$y$ maps with CMB lensing and galaxy lensing maps, respectively. In this paper, we show and compare theoretical predictions for these cross-correlations using both an analytic halo model and full cosmological hydrodynamic simulations that include sub-grid models for radiative cooling, star formation, and AGN feedback. We predict signals for both CMB lensing, $\yphi$, and galaxy lensing, $\ygal$.

Using the gas pressure profile derived from the simulations, we self-consistently compare the halo model predictions to the simulations. The predicted signals from the halo model and simulations agree well over a wide range of angular scales for $\yphi$ and $\ygal$. Small differences are seen at low-$\ell$ where the halo model does not capture the signal from diffuse gas in the intergalactic medium, an effect that is stronger in $\yphi$. However, the diffuse signal comprises only a small fraction of the total signal, even at low-$\ell$. Additionally, at high-$\ell$, the $\yphi$ predictions from the halo model have more power than the simulations, which is a result of the cuspy NFW density profile assumed in the halo model compared to the flatter interior density profile seen in the simulations.

Both $\yphi$ and $\ygal$ are functions of the assumed ICM physics model. However, the ICM models affect $\yphi$ and $\ygal$ differently since the different lensing kernels lead to sensitivity to different halo masses and different redshift ranges. The $\yphi$ observations receive strong contributions from halos with $M_{500} \lesssim 7.1\times10^{13} \, \rmn{M}_\odot$ and $z \gtrsim 0.9$. The mass and redshift dependences for $\ygal$ depend on the specifics of the galaxy lensing survey. For \emph{CFHTLenS}, $\ygal$ is most sensitive to halo masses between $1.3 \times 10^{14} \, \rmn{M}_\sun < M_{500} < 3.4 \times 10^{14} \, \rmn{M}_\sun$ for $\ell \gtrsim 500$, and redshifts $0.05 \lesssim z \lesssim 0.3$ for $\ell \lesssim 1500$ and $0.3 \lesssim z \lesssim 0.5$ for $\ell \gtrsim 1500$. Thus, combining the $\yphi$ and $\ygal$ measurements provides tomographic information on the correlation of matter and ionized gas.

The cross-spectra $\yphi$ and $\ygal$ are sensitive to cosmological parameters in addition to the ICM model. They both roughly scale as $\sigma_8^6$ and $\Omega_\rmn{M}^2$. We compare our results with the existing tSZ -- lensing cross-correlation measurements. The {\it AGN feedback} model with WMAP9 cosmological parameters provides a good fit to the $\yphiname$ results of \citet{HS2014}, although the $\ygalname$ results of \citet{vWHM2014} qualitatively prefer a lower amplitude, particularly at small scales. Due to degeneracies between the gas physics model and cosmological parameters, it is unclear what role diffuse, unbound gas (missing baryons) might play in either measurement. Moreover, such gas only contributes a small fraction of the total signal. Given current observational and theoretical uncertainties, no robust claim can be made at the present time. Comparing the halo model and simulation calculations indicates that the presence of diffuse gas should be seen most clearly at low-$\ell$ in $\yphi$.

Looking ahead, we forecast the constraints on cosmological and astrophysical parameters obtainable with current and future $y-$maps cross-correlated with CMB and galaxy lensing surveys. We show that the combination of these future cross-spectra measurements will constrain ICM physics parameters to $\approx 5-20$\% percent precision, even after marginalizing over cosmological parameters (with the inclusion of primary CMB data). Thermal SZ -- gravitational lensing cross-correlations thus hold immense promise for understanding the physics governing hot, ionized gas throughout the history of structure formation in our Universe.

\acknowledgments

We thank J.R.~Bond, J.~Liu, B.D.~Sherwin, D.N.~Spergel, and L.~van~Waerbeke for useful discussions. We are also grateful to B.D.~Sherwin and R.~Allison for providing ACTPol and AdvACT CMB lensing forecasts on behalf of the ACT collaboration, and to J.~Liu for guidance on the \emph{CFHTLenS} lensing convergence calculations. Before submitting this manuscript, we became aware of similar calculations presented in~\citet{Hojjati2014}, and we subsequently exchanged correspondence with these authors.

\bibliography{bibtex/nab}
\bibliographystyle{apj}

\end{document}